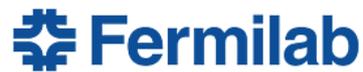

# Studying Qubit Interactions with Multimode Cavities Using QuTiP

**Soumya Shreeram**

The University of Manchester, UK

July-August 2019

This project was done as part of the

**Helen Edwards Summer Internship Program**,

Applied Physics and Superconducting Technology Division

Fermilab, Batavia IL 60510

This project was supervised by **_Yu-Chiu Chao_**,

with support from _Eric Holland_, and was reported to _Alexander Romanenko_.

# Abstract

The project uses QuTiP, a quantum computing framework [1], to simulate interactions between two-qubits coupled with each other via three resonators. The main aim of this project is to build machinery of techniques to understand complex qubit-cavity interactions using QuTiP's functionalities. The system simulated mimics the one constructed by McKay et. al. [2] (M15) and the results of the simulations closely agree with M15's experimental results. The effect of the *coupling strength* between the qubits and the cavities is studied. It was observed that stronger couplings generated larger separations between the eigen-modes. Studies involving *resonance* were used to construct the iSWAP gate, a universal quantum logic gate. This study showed the importance of external thermal losses due to cavity dissipation, and qubit decay and dephasing. The Landau-Zener model was tested for the case of multiple crossing; the model motivated the comparison of three scenarios where the 2 qubits were coupled via 1 cavity, 3 cavities, and 6 cavities. The study concluded that having multiple modes, which is a consequence of having multiple cavities, is advantageous for transferring energy from the qubit to the cavity. Finally, the ac-Stark shift was measured in the system and its dynamics showed excellent agreement with the experimental results obtained by M15.

# Contents

# List of Figures

# 1   Introduction

Computer technology has experienced a paradigm shift in the last few decades, from gears and valves to present-day state of the art transistors, silicon chips, and complex integrated circuitry. The pursuit to achieve smaller computer sizes and more efficient computational techniques has been relentless. The downfall with transistors becoming smaller and smaller is the inescapable dawn of the quantum regime [3]. Richard Feynman, in his 'keynote speech' in 1981 [4], proposed the need for a quantum computer through which it might be possible to simulate quantum mechanical systems and simultaneously enhance computational abilities.

The fundamental unit of information in a classical computer is a bit; it is a two state system represented by a transistor that can be 0 (low) or 1 (high) [3]. The analogue in a quantum computer is a qubit that can be physically represented by any two level system like: ion traps, nuclear and electron spins, photons and superconducting circuits [5, p. 20]. In this project we are interested in the latter, a superconducting qubit called the *transmon*, described in Sec 2.1. In order to build integrated quantum circuits, the entangled qubit states need to be manipulated using high-Q cavities. These manipulations must be coherent such that they minimally contribute to the spontaneous decay or dephasing of the qubit. The study of interactions between superconducting circuits (qubits) with the electromagnetic modes inside a cavity is an extensive field of research that is coined under the term circuit Quantum Electrodynamics (QED) [5, 6]. These cavities can be LC resonators described as a quantum harmonic oscillator [7]. Recent studies have proposed 3D superconducting resonators that enable the achievement of high coherence times (2 sec) at low temperatures $T \leq 20$mK [8].

Finally, to build a quantum computer, one would need quantum logic gates and quantum algorithms. David Deustch and Shor were among the pioneers to formulate quantum algorithms [9, 10], and ever since then the progress in the field has escalated. In January 2019, IBM introduced Q System One which a 20-qubit processor that can be accessed via cloud and is designed for commercial use [11]. In this project, we will be using *QuTiP* [1], a quantum toolbox in Python, that allows us to simulate quantum systems and study their dynamics. The system simulated contains two-qubits coupled via a three-mode filter (LC resonators/ cavities), as shown in Fig. 1.1. The two qubits used are flux-tunable transmons (see Sec. 2.1) and the three-mode filters correspond to a capacitively coupled chain of LC resonators at a frequency $\nu_F$ [2].

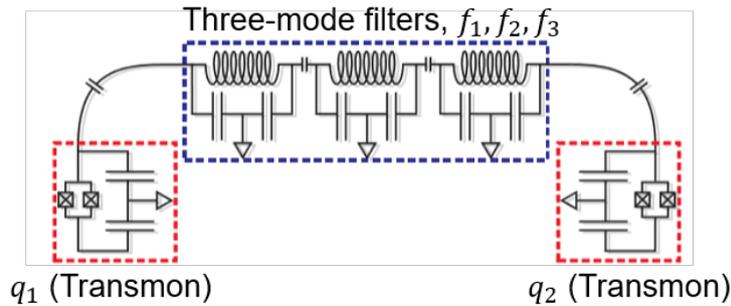

**Figure 1.1:** Schematic circuit diagram of the two-qubit, $q_1$, $q_2$, system which is coupled by the three-mode filters, $f_1$, $f_2$, and $f_3$. Figure adapted from McKay et. al [2].

Sec. 2 covers some of the theoretical concepts required to interpret the results. Sec. 3 and 4 demonstrates the simulation of spectroscopic and resonance studies for the system. Sec. 5 and 6 discusses the dynamics of adiabatic traversing of the cavity-modes of the system, and observing the Stark shift. Finally, the significance of all the simulations using QuTiP is summarized in Sec. 7.

# 2    Theoretical preliminaries

## 2.1    Transmon: A Superconducting Qubit

The Bardeen-Cooper-Schrieffer (BCS) theory describes superconductivity as a macroscopic phenomena due to the formation of a *Cooper-pair* condensate that are similar to a boson-like condensate [12]. Cooper-pairs are a bound state formed by a pair of electrons with opposite spins, in a lattice with energy lower than the Fermi energy. This bound state is formed because the attraction between the electron and the lattice phonons is greater than the Coulomb repulsion between the electrons themselves, so called the electron-phonon effect. This condition results in the formation of a state that satisfies properties required for superconductivity [13]. Before proceeding to understand how this Cooper-pairs can be used to build a qubit, we will familiarize ourselves with the Josephson effect.

*Josephson tunneling* is the tunneling of a Cooper-pair across an insulator that is placed between two superconducting metals. This phenomena displays zero resistance and the insulator used must be a few Å thick [14]. The effect was initially measured by Anderson and Rowell in 1963 [15] where they used a tin oxide barrier between superconducting Sn and Pb.

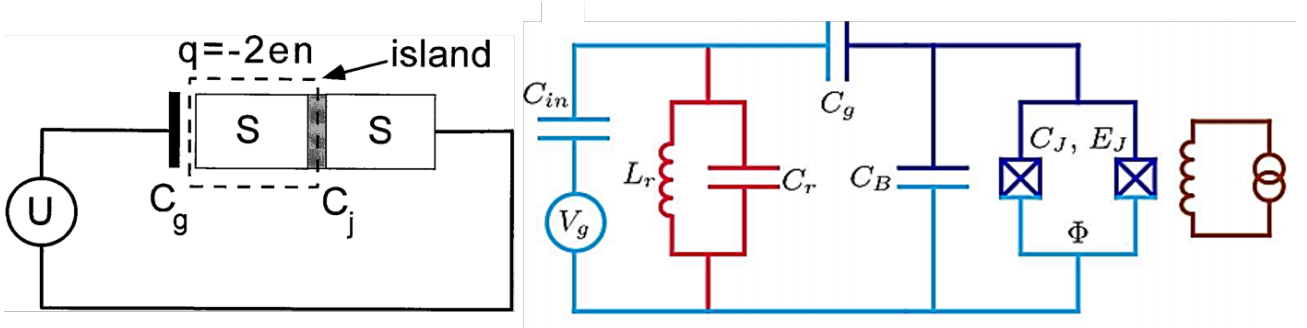

**Figure 2.1:** *(Left)* The Cooper-Pair Box (CBP) containing a tunneling junction (grey zone) between the superconducting "island" and a reservoir with capacitance $C_j$. The excess Cooper-pairs tunnel into the island due to the electric field applied by the voltage $U$ with gate capacitance $C_g$. The schematic and caption is adapted from Bouchiat et. al [16]. *(Right)* The circuit diagram of the *transmon* containing two Josephson junctions (and CBP's) denoted by the checked boxes with energy $E_j$ and capacitance $C_j$. The system is shunted by larger capacitance's $C_B$ and $C_g$. Figure and caption adapted from Koch et. al [17]

The superconducting ring, so called the Cooper-Pair Box (CBP), as demonstrated in Fig. 2.1 (*left*), can mimic a two-level quantum system split by Josephson energy $E_j$; this makes it a desirable candidate for a qubit [16, 18]. Koch et. al (2007) [17] proposed a new superconducting qubit that was closely related to the CBP, called the transmission-line shunted plasma oscillation qubit *transmon*, shown in Fig. 2.1 (*right*). The transmon demonstrated improved insensitivity to charge noise making it an even better candidate for a qubit.

Arriving back to our system of interest as demonstrated by McKay, the Hamiltonian for a system with two qubits with frequencies $v_{Q,1}$, $v_{Q,2}$, and $n$ mode filter can be described as the sum of the qubit Hamiltonian, $\hat{H}_Q$, the filter Hamiltonian, $\hat{H}_F$, and the qubit-filter coupling Hamiltonian, $\hat{H}_{Q-F}$,

$$\hat{H} = \hat{H}_Q + \hat{H}_F + \hat{H}_{Q-F} \tag{2.1}$$

$$\hat{H}_Q = h \, v_{Q,1} \, \frac{\hat{\sigma}_1^z}{2} + h \, v_{Q,2} \, \frac{\hat{\sigma}_2^z}{2} \tag{2.2}$$

$$\hat{H}_F = \sum_{i=1}^{n} h \, v_F \, \hat{a}_i^\dagger \hat{a}_i + \sum_{i=2}^{n} h \, g_F \, (\hat{a}_i^\dagger \hat{a}_{i-1} + \hat{a}_{i-1}^\dagger \hat{a}_i) \tag{2.3}$$

$$\hat{H}_{Q-F} = h \, g_{Q1,F} \, (\hat{a}_1^\dagger \hat{\sigma}_1^- + \hat{a}_1 \hat{\sigma}_1^+) + h \, g_{Q2,F} \, (\hat{a}_n^\dagger \hat{\sigma}_2^- + \hat{a}_n \hat{\sigma}_2^+) \tag{2.4}$$

where $\hat{\sigma}^{+(-)}$ is the raising and lowering operator for the qubit, $\hat{a}_i$ creates a photon in the $i^{th}$ resonator, $g_F$ is the filter-filter coupling, and $g_{Q,F}$ is the qubit-filter coupling [2].

## 2.2   The Landau-Zener Tunneling Model

The Landau-Zener (LZ) model provides a solution to non-adiabatic crossing of energy levels in a two-level system. The model was independently developed by Landau and Zener in 1932 [19, 20]. Although it was originally developed for describing classical systems, its significance in understanding non-adiabatic transitions is manifold [21]. Fig. 2.2 demonstrates the concept. According to the LZ model, the probability for a transition to leave the initial state unchanged depends on the *velocity*, $v$, at the avoided crossing and the *coupling constant*, $J_i$, between the energy modes. If the transition is slow (adiabatic) the final state after crossing remains unchanged with respect to the initial state. However, for a fast crossing (diabatic) the final state is a mixed state.

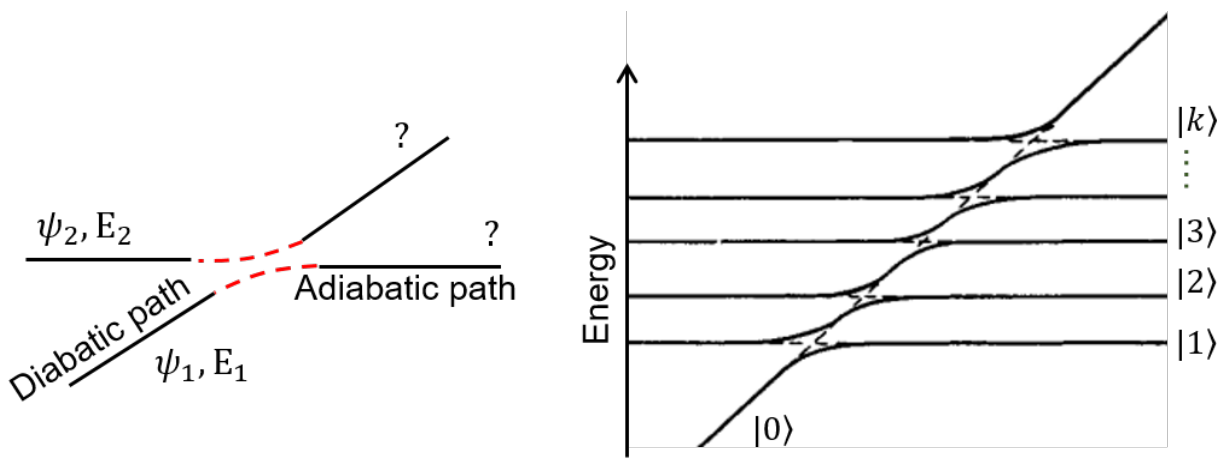

**Figure 2.2:** *(Left)* Schematic showing the crossing of two eigenmodes with energies $E_1$ and $E_2$. The final state after the adiabatic region (highlighted by the dashed-red line) depends on the velocity $v$ with which that crossing occurs and the coupling constant $J_k$ between the eigenmodes. *(Right)* Multiple crossing in the LZ model where the diabatic state $|0\rangle$ undergoes $k$ transitions. Figure adapted from Kayanuma et. al [22].

In particular, we are concerned for the case of non-adiabatic transition in a multiple crossing system. The probability $P$ for state $|0\rangle$ to remain in the same state after the $i^{\text{th}}$ crossing can be given by the product of the probabilities $P_i$ for each independent crossing [22],

$$P = \prod_{i=1}^{k} P_i = \exp\left(-2\pi \sum_{i=1}^{k} J_i^2 |v|^{-1}\right). \tag{2.5}$$

where $i = 1, 2, 3, .., k$ iterates over each independent crossing. Note that the $k$ sub-levels must have sufficient energy gaps to assume independence between subsequent transitions.

## 2.3 Solving the Master Equation

The state of a quantum system can be represented by the density operator $\rho(t)$ for most practical purposes. A closed system that is expressed by the Hamiltonian $H$ follows the von Neumann equation [23, p. 1],

$$i\hbar \frac{d}{dt}\,\rho(t) = [H, \rho(t)] \tag{2.6}$$

which has the solution

$$\rho(t) = U(t, t_0)\rho(t_0)U^\dagger(t, t_0), \text{ where } U(t, t_0) = \exp\left(-\frac{i}{\hbar}H(t - t_0)\right) \tag{2.7}$$

where $\rho(t_0)$ is a suitably chosen initial condition. In a bipartite system, such as a system $\mathcal{A}$ and a reservoir $\mathcal{B}$ ($\mathcal{B} \gg \mathcal{A}$); the evolution of $\rho_\mathcal{B}(t_0)$ is irrelevant. However, the interaction of $\mathcal{A} - \mathcal{B}$ affects $\rho_\mathcal{A}(t_0)$. In majority cases, we are interested in the evolution of $\rho_\mathcal{A}(t_0)$ when taking into account the effects of the reservoir. Such a problem is solved by the Master equation technique which extracts all the desired information relevant for the evolution of $\rho_\mathcal{A}(t)$ from the bipartite system. Kryszewski [23] provides an in-depth tutorial on this subject, which this report will not investigate. In $QuTiP$, the master equation technique is accomplished by `qutip.mesolve` [1]. This interaction of the quantum state with it's environment is accounted by the *collapse operator*, $c_n$, in `qutip.mesolve`. It is represented the second non-Hermitian term in Eqn. 2.8 given as,

$$H_{\text{eff}}(t) = H(t) - \frac{i\hbar}{2}\sum_n c_n^\dagger c_n. \tag{2.8}$$

# 3 Spectroscopy of cavity-cavity and qubit-cavity interaction

A system of multiple coupled cavities at a degenerate frequency undergo splitting when they are coupled, as demonstrated in Fig. 3.1. This effect is well-studied as it has numerous implications in classical as well as quantum systems [24]. The split energy levels are the eigenfrequecies of the filter Hamiltonian $H_\text{F}$ that was introduced in Eqn. 2.3.

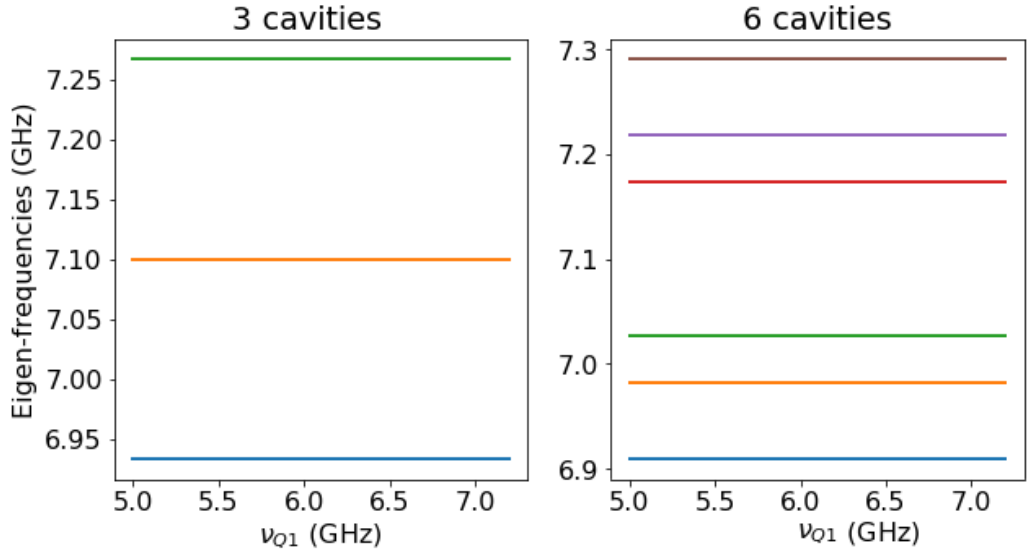

**Figure 3.1:** Spectroscopy of a multi-cavity system for $g_F = 0.135$ GHz. *Left*: A three-cavity system is split into three eigenfrequency due to non-zero coupling between the cavities. *Right*: A six-cavity system is split into six eigenfrequency levels due to coupling between adjacent cavities.

We further explore the effect of coupling in the system when the two qubits $q_1$, $q_2$ are coupled by 3-filter modes $f_1$, $f_2$, $f_3$, shown in Fig. 3.2.

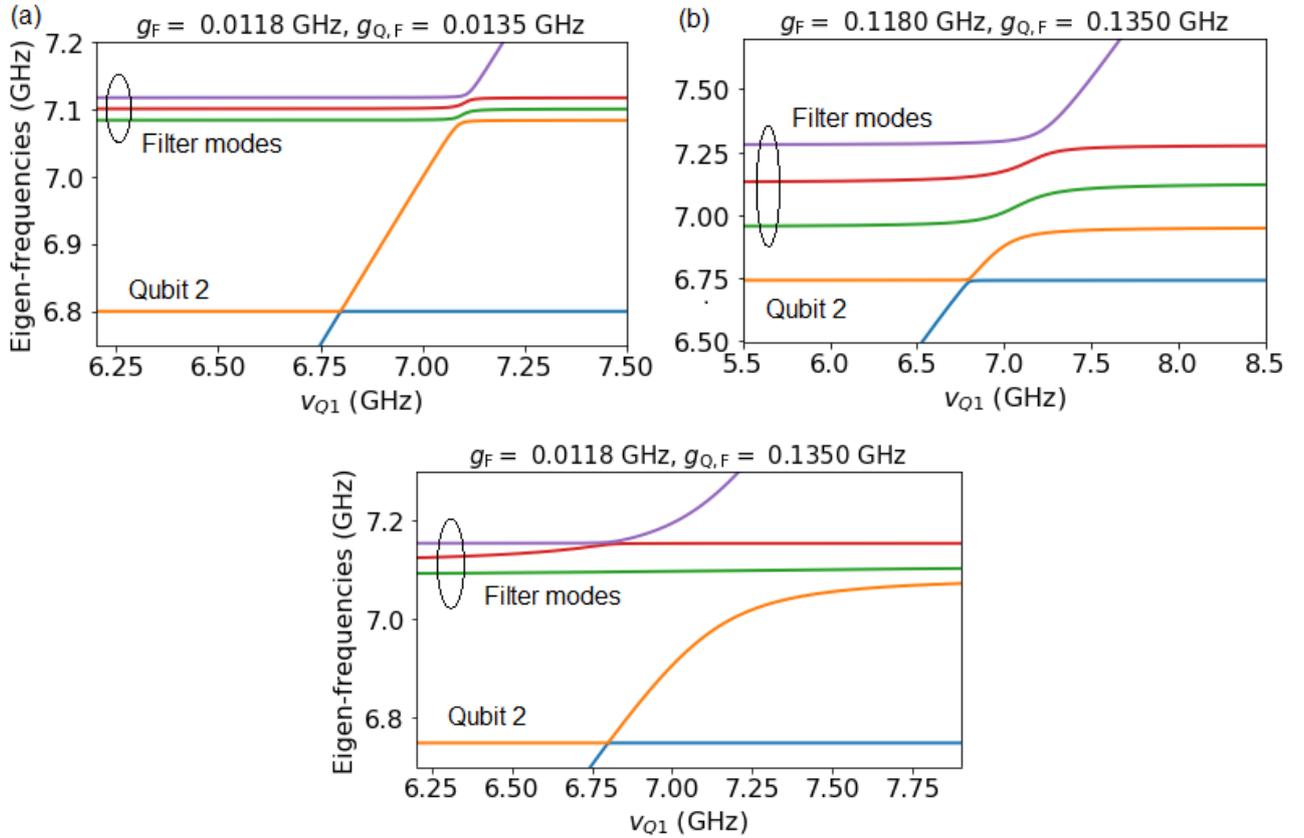

**Figure 3.2:** Effect of coupling strength's on the energy-level splitting in a three-cavity and two-qubit system for (a) low-coupling strength, (b) high-coupling strength, and (c) low-coupling strength between filter modes and high-coupling strength between the qubit and filters.

The study notes some important observations; firstly, the degeneracy in the eigen-frequencies is lifted due to the coupling between adjacent components of the system. Secondly, the distance at the nodes where the degeneracy is lifted depends on the coupling between the components that are interacting at that node. For example, in Fig. 3.2 (a), the distance at the nodes is relatively smaller than for the case (b) where the coupling between the components is 10 times larger. The same point is reiterated in the case (c) where the picture changes because the coupling between the filters is low while the coupling between the $q_1 - f_1$ and $f_3 - q_3$ is high.

# 4    Solving the Master Equation and studying resonance

In this section the evolution of the two-qubit and three-cavity system is studied by solving the Lindblad Master equation by inputting the time-dependent Hamiltonian $H(t)$ into `qutip.mesolve`. Although $H(t)$ can be implemented into `qutip.mesolve` using multiple ways [1], for the following study $H(t)$ is formulated using the *Function-Based* approach [25]. As the three-filters (cavities) originally at frequency $\nu_F$ are coupled, they are split into 3-modes: $\nu_{F1}$, $\nu_{F1}$, and $\nu_{F3}$. Resonance is implemented by raising the qubit frequency $\nu_{Q1}/\nu_{Q2}$ to the 2nd filter-mode at frequency, $\nu_{F2}$.

## 4.1    Applying the iSWAP gate

An iSWAP gate is integral to complete the set of universal quantum gates [26]. The gate is implemented by first bringing $q_1$ in resonance with the 2nd cavity mode, $f_2$, for time, $t_1$. Later, $q_2$ is brought to resonate with $f_2$ for time, $t_2$. The energy is transferred from $q_1$ to the three filter modes $f_1$, $f_2$, $f_3$ at $t_1$. This energy is later transferred to $q_2$ at the end of $t_2$. The effect of the gate applied on the system is recorded by simulating the expectation values of the qubits and cavities in the system as a function of the evolution time, as shown in Fig. 4.1.

The value of $t_1$ is so chosen that $q_1$ completes quarter of a Rabi cycle when it is in resonance with $f_2$. On the contrary, time $t_2$ is allowing for completion of almost half of the Rabi-cycle to occur when $q_2$ is at resonance with $f_2$. This produces a final state where the occupation probability $n_{q_1}(t) \approx 0$ and $n_{q_1}(t) \approx 0.8$; thus, $q_2$ is excited. Thus, they form a crude representation of an iSWAP gate.

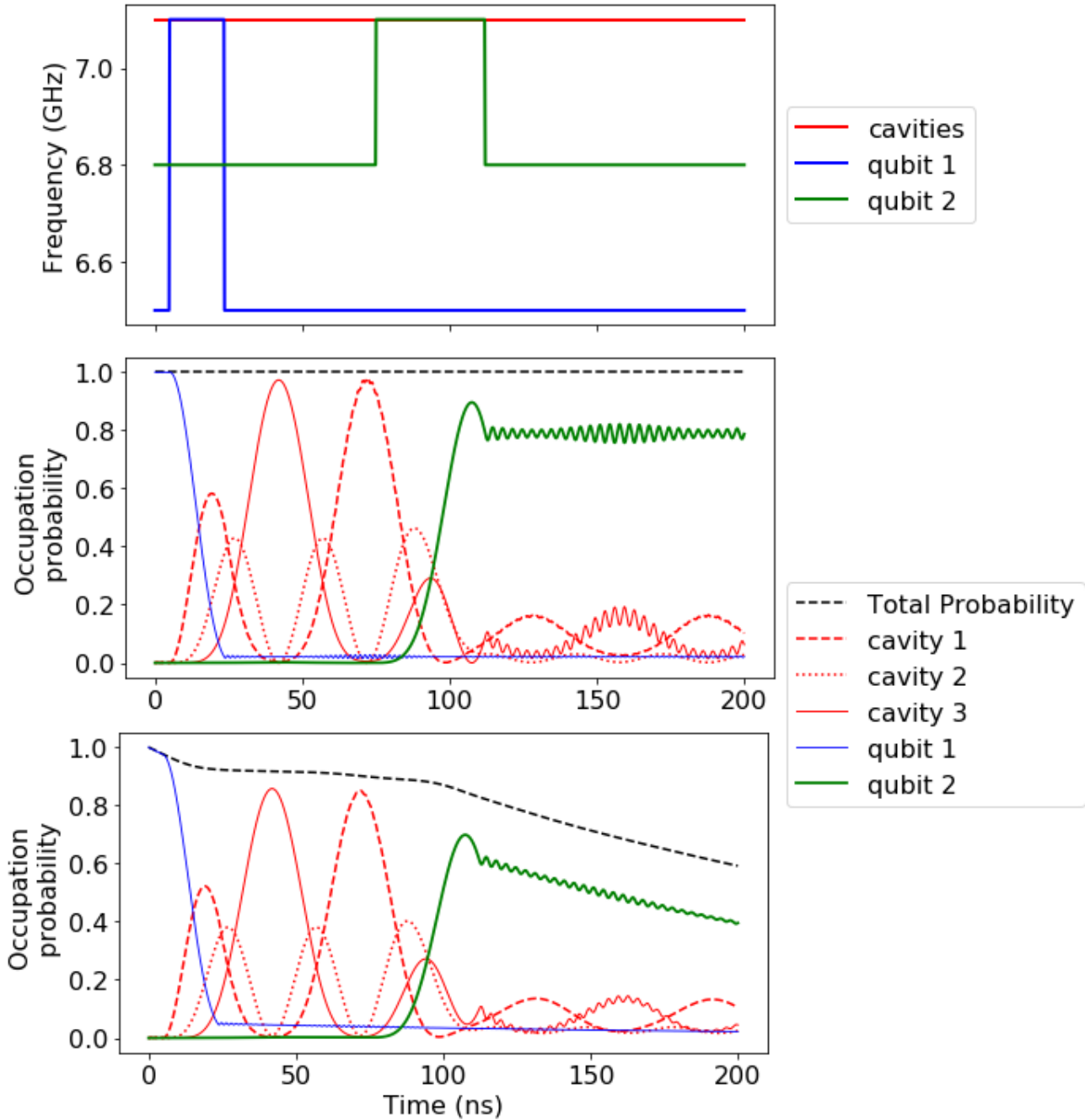

**Figure 4.1:** Implementation of the iSWAP gate on the system. *Top*: The frequncy of $q_1$ is brought to resonate with the second mode of the filter, $f_2$, for time, $t_1 = \pi/\mathbf{2}g_{q,f}$, from 5 ns to $(5 + t_1)$ ns. Similarly, $q_2$ is brought to resonate with $f_2$ for time, $t_2 = \pi/g_{q,f}$, from 75 ns to $75 + t_2$ ns. The coupling between $q_{1,2}$ and $f_2$ was set to, $g_{q,f}/2\pi = 0.0135$ GHz. *Middle & Bottom*: The expectation values as a function of the evolution time without thermal losses (*middle*) and with thermal losses (*bottom*) respectively. See main text for explanation of the resulting behaviour.

Finally, the effect of thermal losses is accounted for when solving the Lindblad Master equation via the collapse operator (see Eqn. 2.8). The leakage or absorption in the cavity is accounted by the photon decay rate, $\kappa$, which was set to 0.1%. The two qubits additionally contributed to a radiative decay rate and dephasing rate of 0.5%. Without accounting for thermal losses, the fidelity and concurrence were simulated to be 72.6% and 25.7% respectively. On considering thermal losses, the fidelity impaired to 45.9%.

## 4.2 Bringing a qubit in resonance with the cavity at finite ramp times

The effect of bringing $q_1$ to resonate with $f_2$ with finite ramp time, $\Delta t$, and for a variable hold times, $t_H$ is studied. The hold time $t_H$ corresponds to the duration for which the pulse is held at the same frequency as $f_2$. Fig. 4.2 shows the effect of changing $t_H$ on the final expectation value achieved by $q_1$ when it is brought off-resonance. The study alludes to two observations:

1. The *final occupation probability of* $q_1$ depends strongly on $t_H$ of the resonant pulse.

2. Fig. 4.2 shows *oscillations in the occupation probability* before it settles to a constant value. These oscillations last for the time corresponding to the $t_H$ of the pulse. They exist because of non-zero coupling strengths, $g_F = 0.0118$ GHz, and $g_{Q,F} = 0.0135$ GHz; the coupling produces the resonant system to Rabi-oscillate.

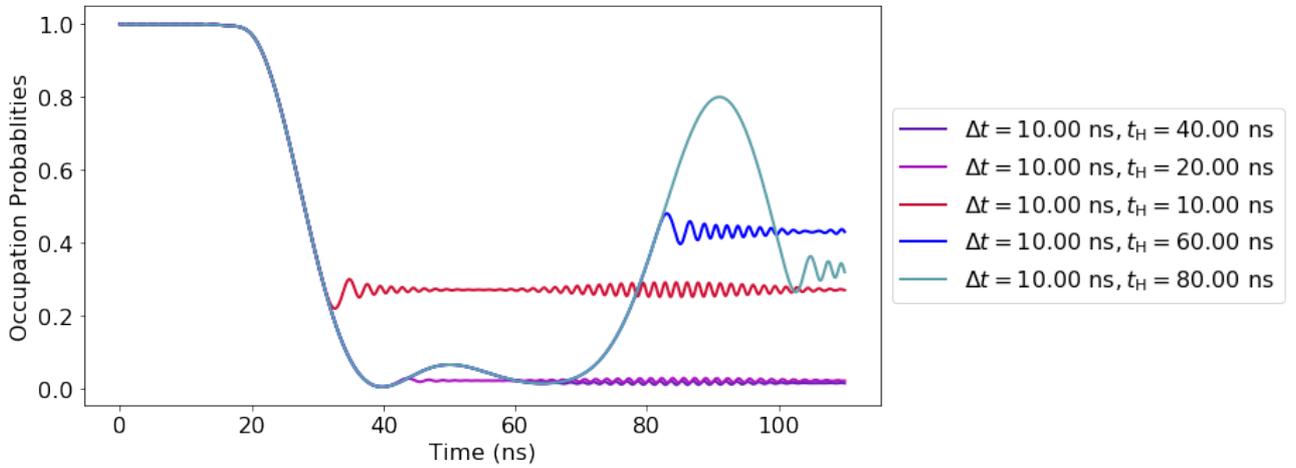

**Figure 4.2:** Occupation probability of $q_1$ as a function of evolution time, $t$, is simulated for a 2 qubit and 3 cavity system. Here, $q_1$ is brought to resonate with the 2$^{\text{nd}}$ mode of the cavity for hold time $t_H$ and with a finite ramp up time, $\Delta t = 10$ ns

This exercise builds towards Sec. 5, where the pulse generated to resonate $q_1$ with the cavity mode is used to study the effect of avoided crossing in the system.

## 4.3   Effect of resonance when applying an RF pulse

A qubit coupled to a cavity can be studied using a tunable probe like a monochromatic field at frequency $\nu$. This realization was investigated by Haroche, and is explained in his book on Cavity Quantum Electrodynamics [27, p. 853]. Work was conducted to explore this subject by perturbing $q_1$ in the following way [27, p. 853],

$$\mathcal{W}_{Q1}(t) = -\hbar\Omega_p[\sigma_1^+ e^{-i\nu t} + \sigma_1^- e^{i\nu t}], \tag{4.1}$$

where $\sigma_1^{+/-}$ are the raising and lowering operators for $q_1$, and $\Omega_p$ is the Rabi frequency of the probe field. The perturbation introduced in Eqn. 4.1 was added to the time-dependent Hamiltonian describing the system. The Lindblad Master equation was solved for the system of 2-qubits that are coupled with 3-cavity modes without accounting for thermal losses (see Eqn. 2.8). The aim of this study was to observe resonance at the eigenmodes of the system Hamiltonian $H(t), \forall\ t \in\ evolution\ times$, as $\nu$ was changed. Although, the task of understanding the appropriate range for the parameters $[\Omega_p,\ \nu]$ remains ongoing and requires further consideration to see plausible results[1].

# 5   Dynamics of adiabatic traversing of qubit-filter avoided crossing

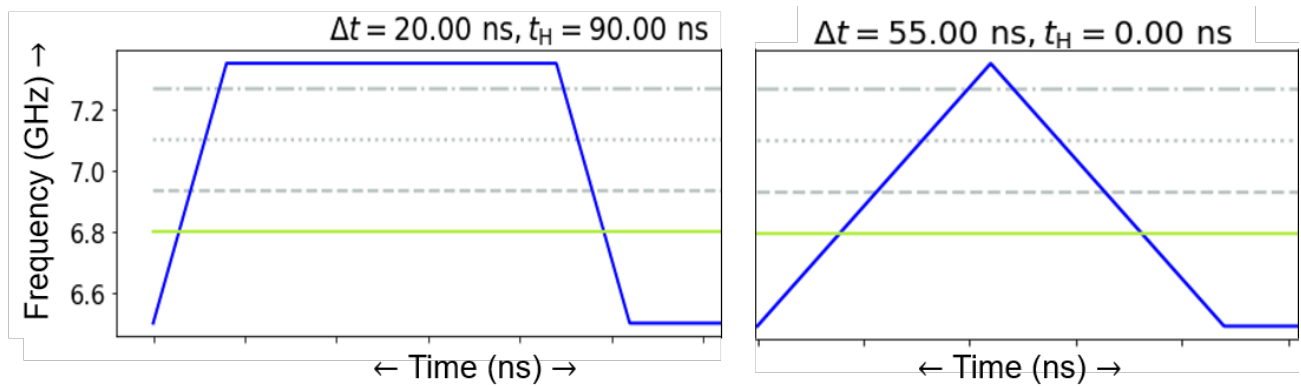

**Figure 5.1:** Demonstration of two types of ramps crossing $q_2$, $f_1$, $f_2$, and $f_3$ modes. The pulse (blue) is inputted into the time-dependent Hamiltonian given by Eqn. 2.8 without accounting for thermal losses. *Left*: The pulse has an adiabatic ramp time, $\Delta t_{ramp} = 20$ ns. *Right*: A *completely* adiabatic ramp for the same total period as the one to the *left*, $T = 110$ ns.

---

[1]Preliminary results are available to view on the GitHub repository: github.com/SoumyaShreeram/Qubit/.

We progress towards understanding the dynamics of non-adiabatic transition after multiple-crossing for our system of interest. The theory was briefly introduced in Sec. 2.2, and the execution of this phenomena is demonstrated in Fig. 5.1. A range of different pulses, with total period $T = 110$ ns, were inputted into the time-dependent Hamiltonian $H(t)$, with different ramp times $\Delta t_{ramp} \in [0, 55]$ ns. The occupation probability, $n(t)$, of $q_1$ was recorded at the end of the pulse i.e. after the ramp-down, $n(t_{\text{down}})$. Fig. 5.2 (*left*) shows the variation in $n(t_{\text{down}})$ as a function of the the ramp-up/down time of the corresponding pulses. This procedure was repeated for three systems, where the two-qubits were coupled via 1 cavity, 3 cavities, and 6 cavities.

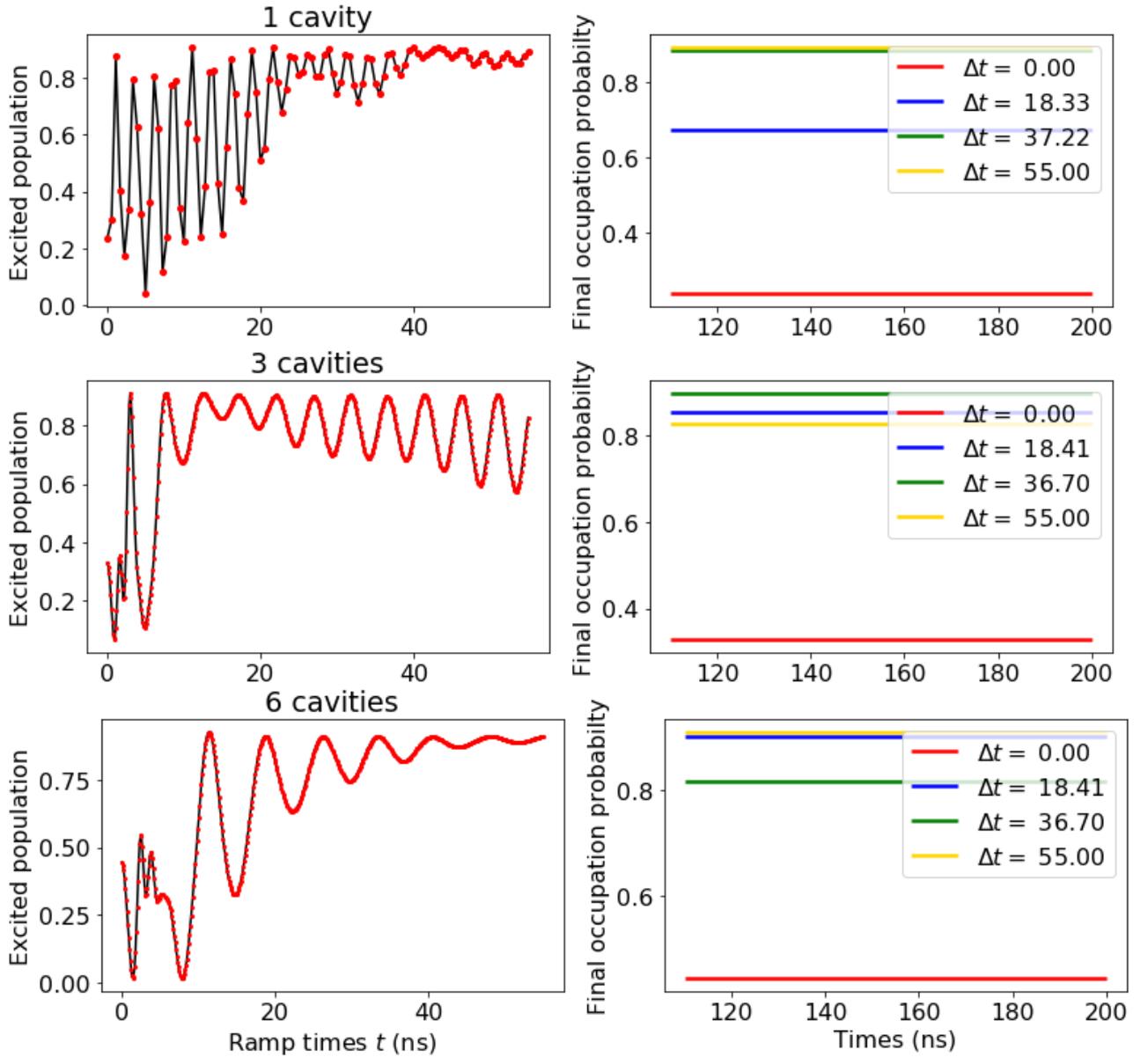

**Figure 5.2:** (*Left*) Fringes observed in the occupation probability of $q_1$, at the end of the pulse, as a function of changing ramp times $t$ for the 1, 3, and 6 cavity cases. (*Right*) The final occupation probability of $q_1$ as a function of evolution time after the ramp-down of the pulse is plotted for four ramp times i.e. they represent a data point (red dot) on the plot to the *left*.

This exercise alludes to the following observations:

1. The Landau-Zener model implies that large ramp times (slower velocities) preserves the final state of $q_1$, whereas, short ramp times (faster velocities) results in a mixed state of $q_1$. By studying the fringe-frequencies that are observed in Fig. 5.2 (*left*) at short ramp times $\lesssim 20$ ns relative to the large ramp times $\gtrsim 20$ ns, higher-amplitude and higher-frequency oscillations are noted for $t_{\mathrm{ramp}} \lesssim 20$ ns compared to the large ramp times. This agrees with result we would expect from the LZ model. However, an anomaly was observed for the 3 cavity case, where the oscillations are noted to grow slightly in amplitude with the increase in ramp times. This could be a numerical-error artifact, however, this claim requires further study to hold true as it was not observed in the 1 cavity and 6 cavity case.

2. Following the previous qualitative discussion of the fringe frequencies for the 1, 3, and 6 cavity cases, the same is proved quantitatively by taking the fast Fourier transform of the plots in Fig. 5.2 (*left*) for short ramp times $\lesssim 20$ ns and at large ramp times $\gtrsim 20$ ns, as shown in Fig. 5.3. The comparison of the frequency peaks for the 1, 3, and 6 cavity cases for $t_{\mathrm{ramp}} \lesssim 20$ and $t_{\mathrm{ramp}} \gtrsim 20$ show a common trend that is, *increasing the filter-modes (cavities), reduces the fringe frequency.*

3. McKay et. al (M15) [2] conducted a similar study experimentally for the 3-cavity case; the result obtained by them closely resembles those simulated and shown in Fig. 5.2. M15 also claimed that the multi-mode nature is advantageous for adiabatically traversing filter modes i.e. it is easier to load a photon (transfer energy) from $q_1$ to $f_1$ at smaller ramp times for a 6 cavity case relative to the 1 cavity case. We ignore the 3 cavity case while making this comparison due to the anomalies in its results that are not yet understood. Nevertheless, the claim that the multi-mode nature is advantageous is successfully shown when comparing the 1 and 6 cavity case, see Appendix Fig. 7.1.

4. Finally, the plots to the right of Fig. 5.2 emphasize the final state occupation probability of $q_1$, i.e., the correspond to $n(t) \, \forall \, t > t_{down}$. Notice, that the 0 ramp time scenario (fast velocity) results in inefficient transfer of energy between qubit and cavities. However, larger ramp times show that $q_1$ efficiently loads and retrieves the energy from the cavities.

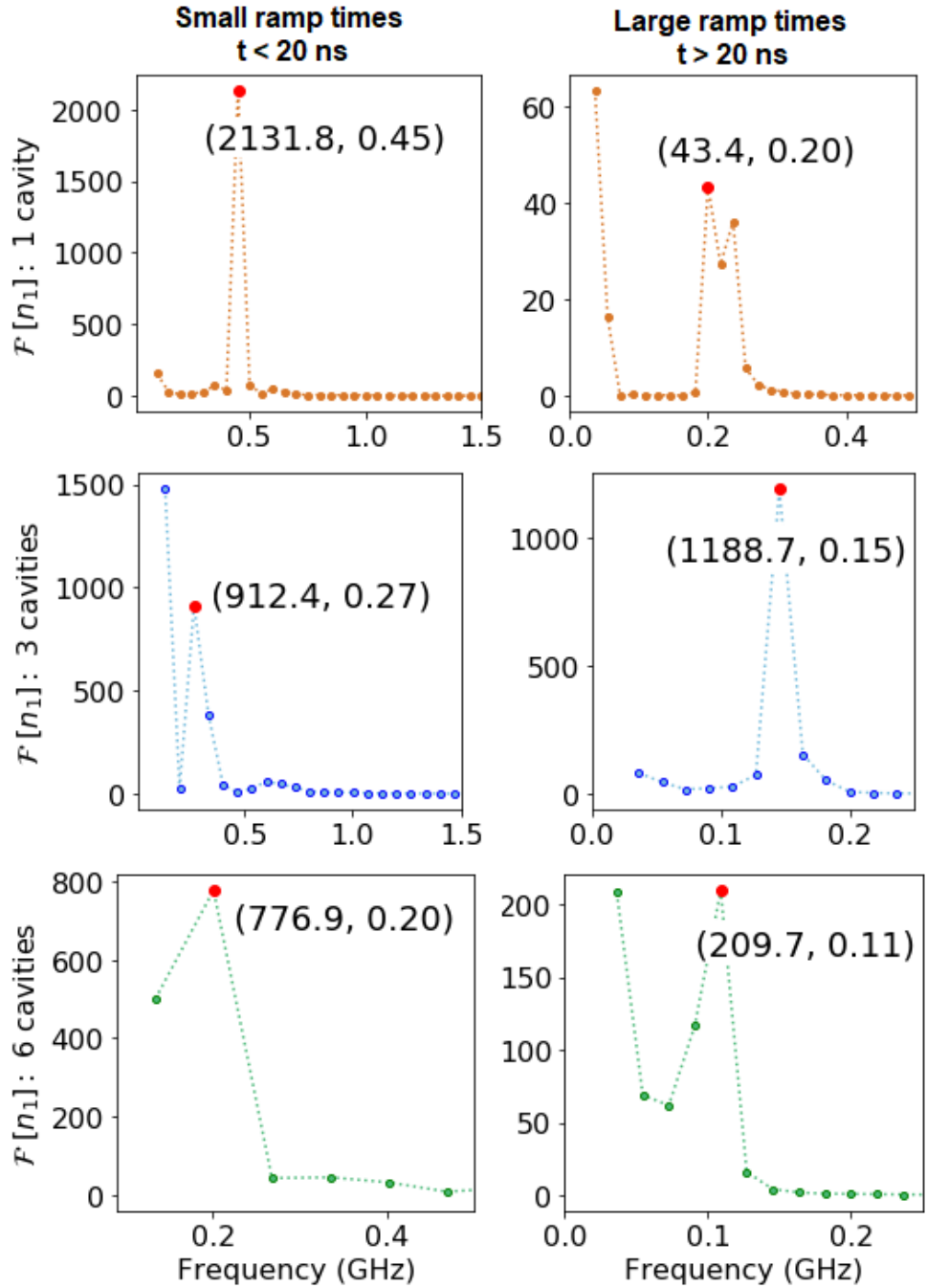

**Figure 5.3:** The Fourier transform of the plots in Fig. 5.2 (*left*) for short ramp times $\lesssim 20$ ns and large ramp times $\gtrsim$ ns.

# 6 Stark shift

We must note that the important parameters in circuit QED are the cavity resonance frequencies $\omega_F = 2\pi\nu_F$, the qubit-transition frequencies $\Omega_{q_1/q_2} = 2\pi\nu_{q_1/q_2}$, and the coupling frequency between the cavity modes and the qubits $g_{Q-F}$, which were observed in Eqn. 2.1. This leads us to define the qubit-cavity detuning, $\Delta = \omega_F - \Omega_{q_1/q_2}$, which is used to recognize two limiting cases:  large detuning when $g_{Q-F}/\Delta << 1$, and no detuning when $\Delta = 0$.  Sec. 3 and 4 were discussing results in the latter limiting case, however, Sec. 5 and 6 are concerned with the former limiting case.  Specifically, in the large detuning limit, the qubit pulls the cavity frequencies by $\pm g_{Q-F}^2/\Delta$.  This is a phenomena understood as the ac-Stark/Lamb shift [28].  The methodology followed for calculating the stark shift is demonstrated in Fig. 6.1 and 6.2.  It can be summarized in 4 steps:

1. The initial state of $q_1$ was prepared as a superposition $(|0\rangle + |1\rangle)/\sqrt{2}$, and $q_2$ was initially in the ground state.

2. $q_1$ was excited above the cavity modes for a fixed period $t_H = 90$ ns with a ramp time of $\Delta t_{\text{ramp}} = 10$ ns.

3. $q_2$ was excited to a variable height (frequency) $\nu_{Q2}$, and a series of 500 pulses each with a different pulse width $\tau$ were constructed.

4. A $\pi/2$ pulse was applied after $q_1$ was brought to its initial frequency (off-resonance).  The entire system, with the variations in $q_1$ and $q_2$, were solved for using the time-dependent Lindblad master equation (see Eqns. 2.6, 2.8).  The occupation probability of $q_1$, $n_1(t)$, was recorded after the $pi/2$ pulse was applied.

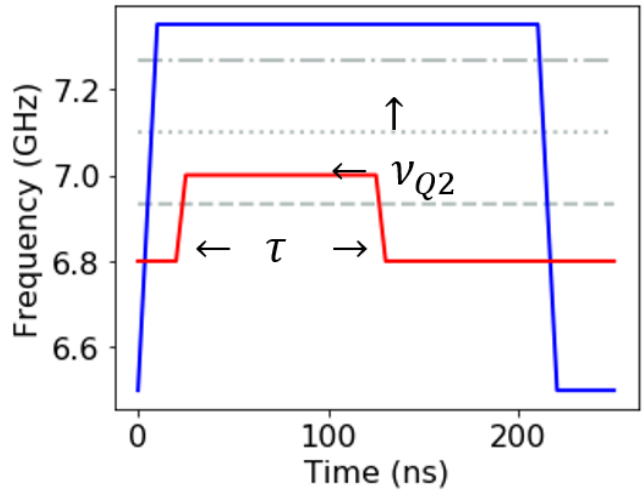

**Figure 6.1:** Demonstration of the set-up to measure the Stark shift in a 2-qubit, 3-cavity system. The blue line represents the variation in the frequency of $q_1$, whereas, the red line is the variations in the frequency of $q_2$. The three-grey lines are the three coupled cavity modes. $\nu_{Q2}$ represents the height upto which $q_2$ is raised and $\tau$ is the parameter that executes variations in the pulse width of $q_2$.

For various $\tau$, and fixed $\nu_{Q2}$, the Ramsey fringes were observed due to the variable interaction times, as shown in Fig. 6.2 (*left*). The frequency of these fringes was measured by implementing the Fourier transform technique, as shown in Fig. 6.2 (*right*).

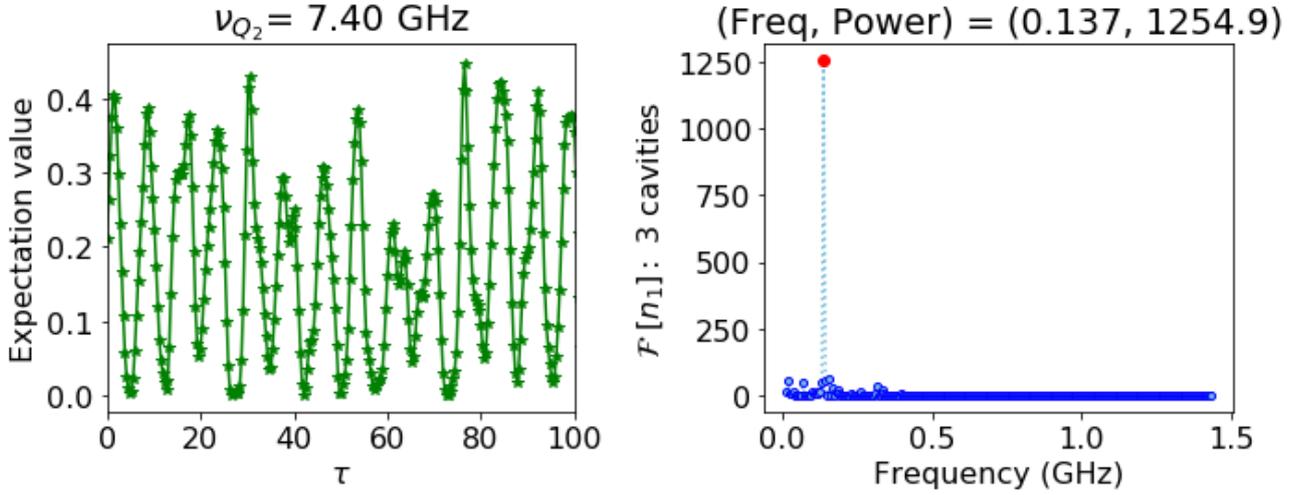

**Figure 6.2:** Ramsey fringes and their Fourier transform. (*Left*): The expectation value of $q_1$ recorded after Step 4, as a function of the variable pulse widths $\tau$ (ns) of $q_2$. (*Right*): The Fourier transform of the Ramsey fringes observed to the left.

Each data point shown in Fig. 6.3 (*left*) is obtained by repeating the procedure discussed in steps 1 to 4 for varying $\nu_{q2}$. The Stark shift increases as $\approx 1/\Delta$ as $q_2$ approached the filter from below. The shift-frequencies are observed to saturate at $150MHz \approx \sqrt{2}g_F$, as claimed by M15 [2]. Fig. 6.3 (*right*) is the experimentally obtained data by M15 with a theory curve. The simulated curve (*left*) closely resembles the results obtained by M15, validating their claims.

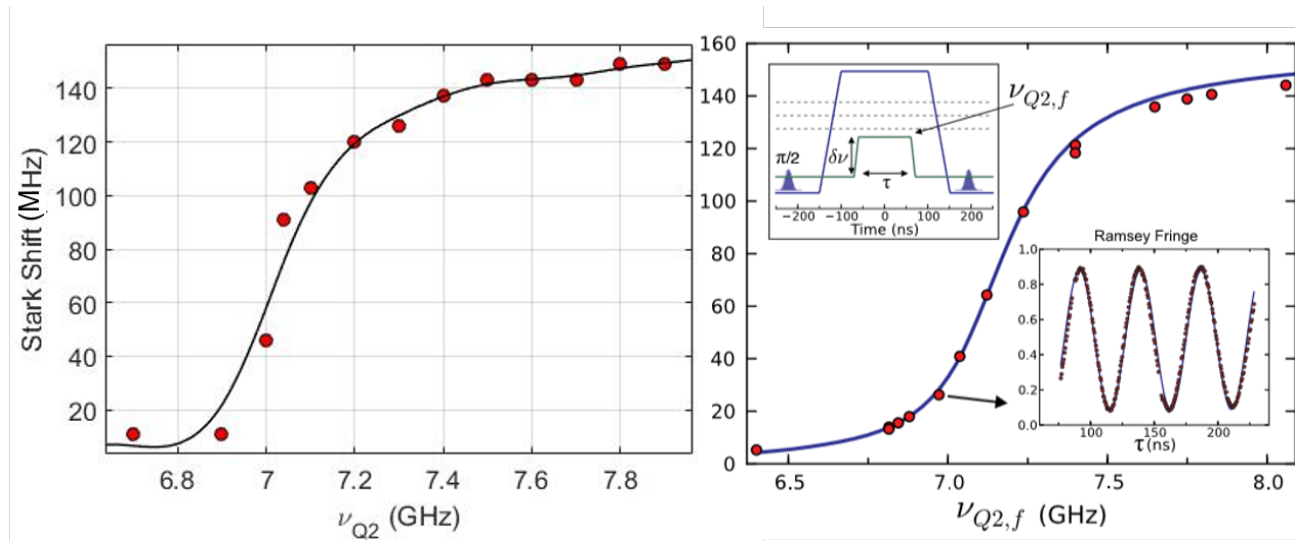

**Figure 6.3:** The Stark shift plotted as a function of the $q_2$ frequency. Upon comparison, the results simulated on QuTiP (*left*) with those obtained by McKay et. al. [2] (*right*) show resemblance.

# 7    Conclusions

The project studies the interaction of a two qubit system coupled with multi-mode cavities. The simulations were aided via QuTiP and the results were obtained by successfully using QuTiP's functionalities like: solving the master equation `qutip.mesolve`, using partial traces `qutip.ptrace`, using the time-dependent Hamiltonian, etc. The eigenstates of the system Hamiltonian were non-degenerate due to coupling terms and the distance between the eigenmodes was strongly dependent on the magnitude of the coupling. By simulating the effect of resonance of the qubits with the cavity modes, the iSWAP gate was constructed and the effect of the thermal channel was accommodated. Finally, the last two sections of this project was motivated by the work conducted by McKay et. al. The dynamics of the multi-mode system was studied by testing the Landau-Zener model and by conducting the Ramsey experiment to observe the ac-Stark shift. The latter two features could be further carried on to build the controlled-Z gate, as formerly shown by McKay.

# Appendix

This section contains some further material that are useful aids to the results discussed in Sec. 6.

**Comparison for adiabatic traversing of filter modes for 1, and 6 cavity cases**

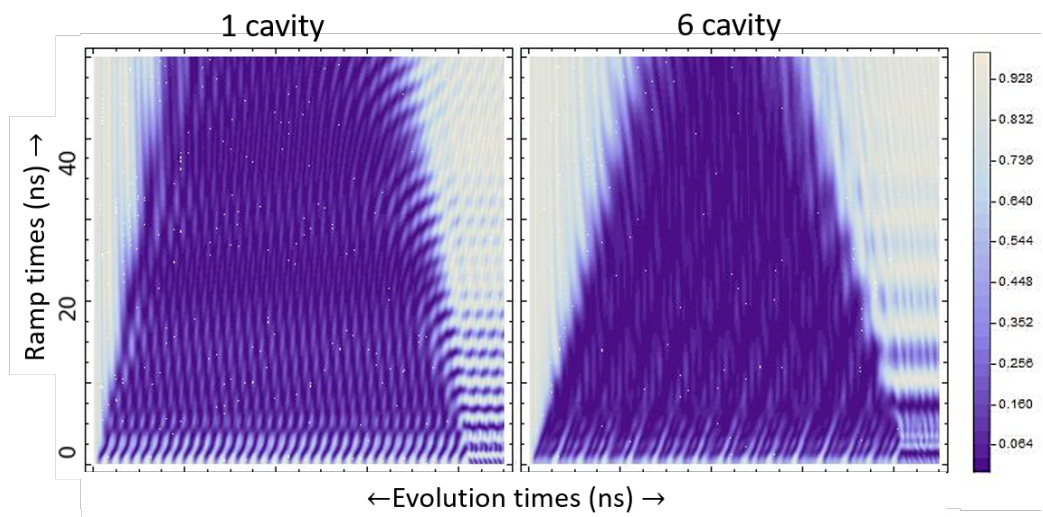

**Figure 7.1:** Contour plots comparing the 1 and 6 cavity case for photon loading (energy transfer).

# Acknowledgements

Special thanks to my supervisor Yu-Chiu Chao for his guidance and patience throughout the 10-weeks of my project at Fermilab. His enthusiasm and innovation provided me with a great learning experience. I would also like to thank Eric Holland for his help with my code at crucial times and Alex Romanenko for his encouragement. Finally, I'm grateful to the Helen Edwards internship committee for providing me with this opportunity.

This document contains 3180 words (without references, captions, headers).